\documentclass[conference]{IEEEtran}
\IEEEoverridecommandlockouts

\usepackage{cite}
\usepackage{amsmath,amssymb,amsfonts}
\usepackage{algorithmic}
\usepackage{graphicx}
\usepackage{textcomp}
\usepackage[dvipsnames]{xcolor}
\def\BibTeX{{\rm B\kern-.05em{\sc i\kern-.025em b}\kern-.08em
    T\kern-.1667em\lower.7ex\hbox{E}\kern-.125emX}}

\usepackage{bbm}
\usepackage{breqn}
\usepackage{amsthm, mathtools}
\usepackage{accents}
\usepackage{hyperref}

\newcommand{\ubar}[1]{\underaccent{\bar}{#1}}

\newtheorem{thm}{Theorem}

\newtheorem{lemma}[thm]{Lemma}

\newtheorem{cor}[thm]{Corollary}
\newtheorem{definition}[thm]{Definition}

\newcommand{\R}{\mathbb{R}}

\newcommand{\norms}[1]{\left\lVert#1\right\rVert}
\newcommand{\norm}[1]{\left | #1\right |}
\DeclareMathOperator*{\argmax}{arg\,max}

\DeclareMathOperator*{\CMI}{CMI}

\newenvironment{sizeddisplay}[1]{\par\nopagebreak#1\noindent\ignorespaces}{\nopagebreak\ignorespacesafterend}

\begin{document}

\title{Potential Contrast: Properties, Equivalences, and Generalization to Multiple Classes\\
\thanks{WP is supported by an EPSRC DTP grant EP/W524633/1. AB and WP thank C2D3 (made possible by Schmidt Sciences) for funding the project \textit{AI meets cultural heritage: non-invasive imaging and machine learning techniques for the reconstruction of degraded historical sheet music}.}
}


\author{\IEEEauthorblockN{Wallace Peaslee}
\IEEEauthorblockA{\textit{DAMTP, University of Cambridge} \\
Cambridge, UK \\
https://orcid.org/0000-0002-5274-9035}
\and
\IEEEauthorblockN{Anna Breger}
\IEEEauthorblockA{\textit{DAMTP, University of Cambridge} \\
Cambridge, UK \\
https://orcid.org/0000-0001-8878-5743}
\and
\IEEEauthorblockN{Carola-Bibiane Schönlieb}
\IEEEauthorblockA{\textit{DAMTP, University of Cambridge} \\
Cambridge, UK \\
https://orcid.org/0000-0003-0099-6306}

}

\maketitle

\begin{abstract}
Potential contrast is typically used as an image quality measure and quantifies the maximal possible contrast between samples from two classes of pixels in an image after an arbitrary grayscale transformation. It has been applied in cultural heritage to evaluate multispectral images using a small number of labeled pixels. In this work, we introduce a normalized version of potential contrast that removes dependence on image format and also prove equalities that enable generalization to more than two classes and to continuous settings.
Finally, we exemplify the utility of multi-class normalized potential contrast through an application to a medieval music manuscript with visible bleedthrough from the back of the page. We share our implementations, based on both original algorithms and our new equalities, including generalization to multiple classes, at \url{https://github.com/wallacepeaslee/Multiple-Class-Normalized-Potential-Contrast}.


\end{abstract}


\section{Introduction} \label{Sec:Introduction}
Potential contrast (PC) is a task-dependent image contrast and quality measure. It involves binarizing an image based on labeled pixels from two classes, typically called the foreground and background, which are usually  selected manually for a particular task. PC then measures the maximal contrast possible between the labeled pixels from each class after an arbitrary grayscale transformation \cite{shaus2017-PC-Math}.

The most prominent successes of PC so far have been in applications to cultural heritage, especially multispectral images of degraded writing. In particular, areas where ink is present (foreground) and absent (background) are labeled. Then PC is computed for each band in a multispectral image using pixels from labeled regions to determine which band(s) may contain the most relevant information. This process is described with more detail in \cite{faigenbaum2012-PC-Method-Ostraca} and \cite{sober2014-PC-Method-Ostraca-Conference}, where PC was  applied to ostraca (potsherds with writing), and further explored in \cite{faigenbaum2022-PC-Example_BITS, faigenbaum2017-PC-Example_Biblical, faigenbaum2014-PC-Example_Hieratic}.

A primary property of PC is its invariance under invertible grayscale transformations, which can account for limitations in human perception like the relative difficulty of evaluating the quality of bright images according to the Weber-Fechner Law \cite{shaus2017-PC-Math, fechner1948_psychology}. For example, an image $I$ may appear to show writing more poorly than another image $J$, when in reality $I$ contains valuable information and shows writing much more clearly after remapping the grayscale values e.g. adjusting brightness or contrast. In that sense, PC does not necessarily correspond to visual perception quality, but rather measures the quality of underlying information. 

\begin{figure}[b!]
\centering
\includegraphics[width=0.24\textwidth]{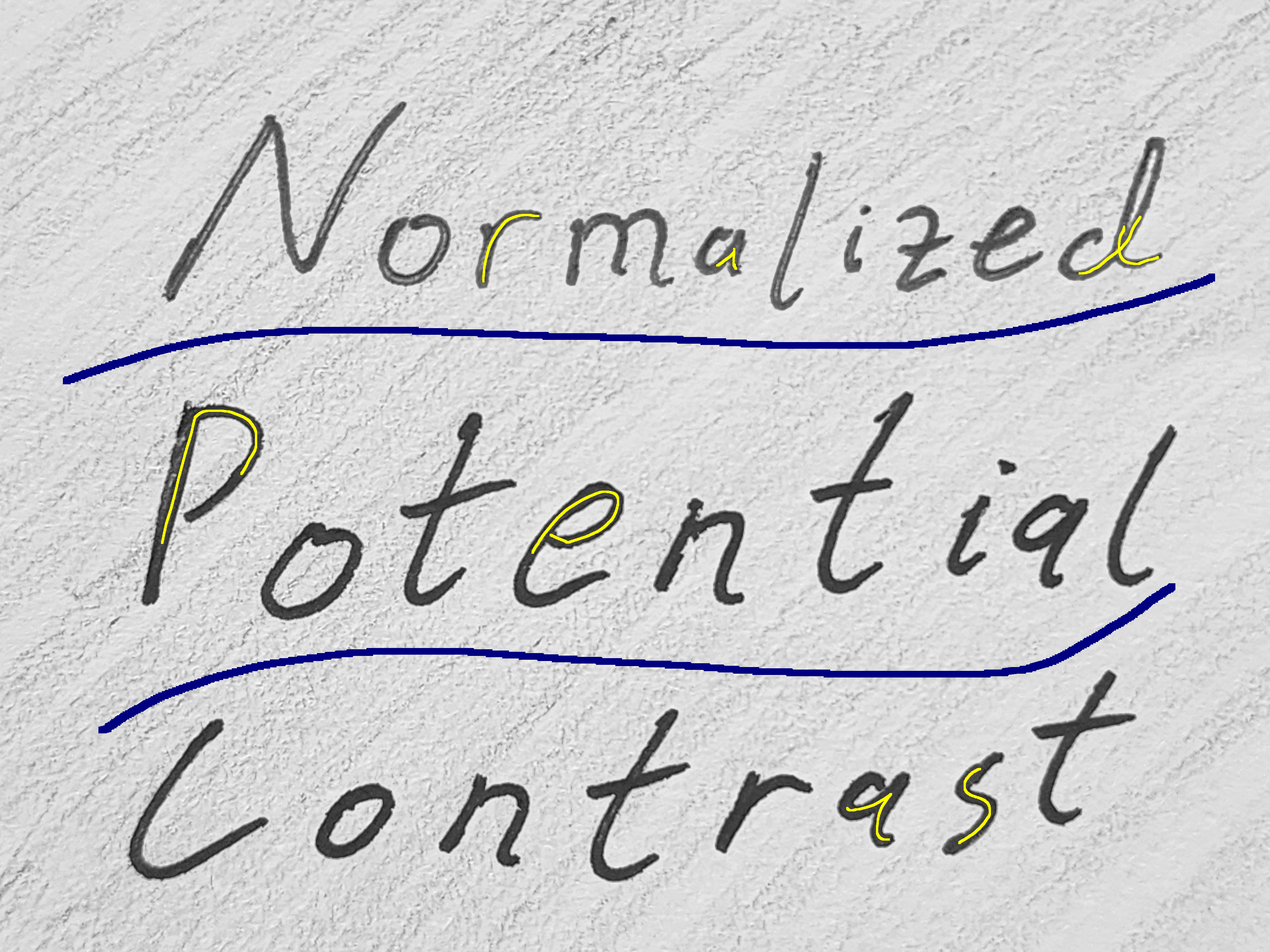}
\includegraphics[width=0.24\textwidth]{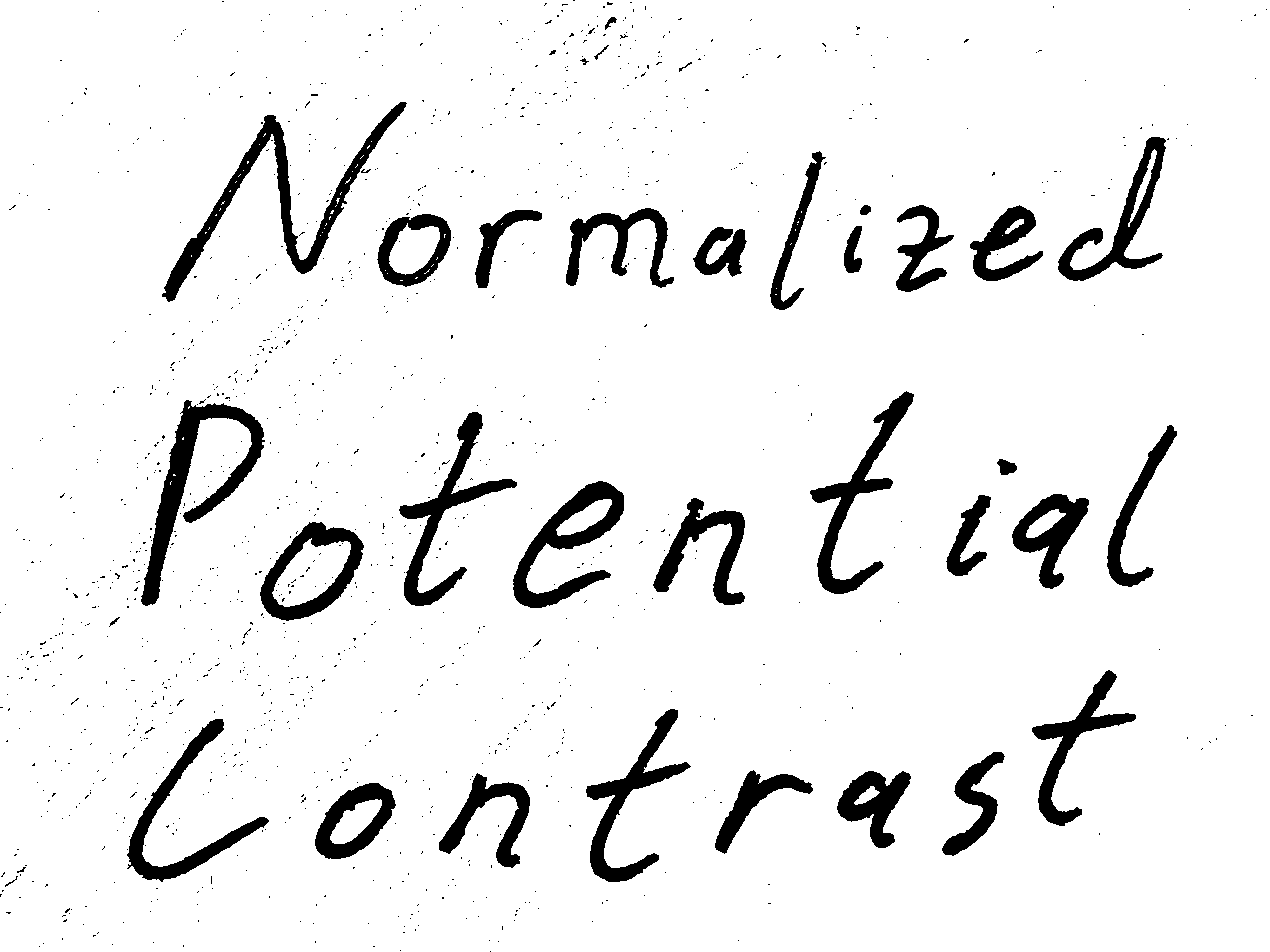}
\caption{(Left) An example image, where labeled background pixels are shown in blue and labeled foreground pixels are in yellow. (Right) The binarization resulting from potential contrast, with a value of 254.983 for an 8-bit image, and a normalized potential contrast value of 0.996.
}
\label{fig: normalized potential contrast}
\end{figure}

In this paper, we introduce the notion of \textit{normalized potential contrast} (NPC), which is based directly on PC, but is commensurate across applications as it does not vary with image format and is interpretable as an accuracy rate of a binary classifier. We prove that NPC is equivalent to the total variation distance \cite{Tsybakov2009_TV} between the distributions of sampled foreground and background pixels, enabling generalization to continuous contexts.

We also introduce new equalities for NPC that can make computation simpler and provide alternate interpretations of its value. One of these equalities allows us to define \textit{multi-class NPC}, generalizing the notion of NPC to more than two classes. Instead of binarizing an image, multi-class NPC segments an image by class. This is useful for several problems arising in cultural heritage including palimpsest, bleed-through, the presence of multiple pigments, and other situations commonly encountered in historical manuscripts. 

Lastly, we provide code for PC, NPC, and their multiple-class generalizations at \url{https://github.com/wallacepeaslee/Multiple-Class-Normalized-Potential-Contrast}. We include implementations following the algorithms described in \cite{faigenbaum2012-PC-Method-Ostraca, sober2014-PC-Method-Ostraca-Conference} as well as implementations using our equalities from Section \ref{Sec:PC-Results}.

This paper is structured as follows. In Section \ref{Sec:PC-Definition}, we define NPC and demonstrate some of its advantages. In Section \ref{Sec:PC-Results}, we introduce equalities that enable generalizations of NPC. In Section \ref{Sec:Multiple Sources}, we define multi-class NPC. Finally, an example application to bleedthrough in historical manuscripts is explored in Section \ref{Sec:Application}.

\section{Potential Contrast \& Normalization} \label{Sec:PC-Definition}
We define potential contrast (PC) based on \cite{shaus2017-PC-Math}. Let $A = (a_1, \ldots, a_n)$ and $B = (b_1, \ldots, b_m)$ be labeled foreground and background pixel values, taken from an image with values from a set $X \subset \R$. We assume throughout this work that $1<|X|<\infty$, i.e. the image is not constant and is finite.  A measure of contrast, `Clayness Minus Inkiness' (CMI), introduced and used to analyze historical documents in \cite{Shaus2012-CMI}, is given by 
\begin{equation*}
   \CMI(A,B) = \mu[A] - \mu[B],
\end{equation*}
where $\mu[A]$ 
denotes the (discrete) mean of sampled foreground pixels and, similarly, $\mu[B]$ denotes the mean of sampled background pixels.  
Additionally, denote the set of functions from a set $X$ to itself by $G(X) = \{g:X \to X\}$. Then, for a given $g \in G(X)$, we define $g(A)$ and $g(B)$ as the sampled foreground and background pixels after applying $g$. Therefore, if $A = (a_1, \ldots, a_n) \in X^n$, then $g(A) = (g(a_1), \ldots , g(a_n)) \in X^n$.
Following \cite{shaus2017-PC-Math}, potential contrast can be defined as
\begin{equation} \label{eqn: potential contrast definition}
    PC_X(A, B) := \max_{g \in G(X)} \CMI(g(A), g(B)).
\end{equation}
Throughout this work, we use $P_A$ to denote the discrete distribution, i.e. the relative histogram of values, in $A$, and likewise for $P_B$. So, $\sum_{x \in X} P_A(x) = \sum_{x \in X} P_B(x) = 1$ and $0 \leq P_A(x), P_B(x) \leq 1$ for any $x \in X$. A solution for Eq. \ref{eqn: potential contrast definition}, namely an optimal grayscale transformation, is the binarization given in Proposition 1 of \cite{shaus2017-PC-Math}:
\begin{equation} \label{eqn: optimal pc g}
g^{opt}_{A,B}(x) = 
\begin{cases}
    \max(X) \text{ if } P_A (x) \geq P_B(x)\\
    \min(X) \text{ if } P_A (x) < P_B(x).
\end{cases}
\end{equation}
Here, the definition of PC and its solution relies on the set $X$, which corresponds to an image's data type. 
To remove this dependence, we introduce a normalized version of PC.
\begin{definition}
    Let $Y$ be the set of values that occur in either $A$ or $B$. Let $H(A,B) = \{h: Y \to \{0,1\}\}$. Then, the normalized potential contrast is given by
    \begin{equation}
        NPC(A,B) := \max_{h \in H(A,B)} \CMI(h(A), h(B))
    \end{equation}
\end{definition}
Following an argument analogous to that of Proposition 1 of \cite{shaus2017-PC-Math}, an optimal $h^{opt}_{A,B} \in \argmax_{h \in H(A,B)} \CMI(h(A), h(B))$ is the binarization
\begin{equation} \label{eqn: optimal npc g}
h^{opt}_{A,B}(y) = 
\begin{cases}
    1 \text{ if } P_A (y) \geq P_B(y)\\
    0 \text{ if } P_A (y) < P_B(y).
\end{cases}
\end{equation}
The manner in which PC relies on the range of values of an image format is formalized by the following lemma.
\begin{lemma} \label{lemma: PC scaling with S}
    For any injective function $f: X \to T \subset \R$ (with $T$ finite),
    \begin{equation*}
        PC_{T}(f(A), f(B)) = \frac{\max(T) - \min(T)}{\max(X) - \min(X)}PC_X(A,B).
    \end{equation*}
\end{lemma}
\begin{proof}
For brevity, let $\bar{x} = \max(X)$, $\ubar{x} = \min(X)$, $\bar{t} = \max(T)$, and $\ubar{t} = \min(T)$. Additionally, let $g_1 \in \argmax_{g \in G(X)} \CMI (g(A), g(B))$ and $g_2 \in \argmax_{g \in G(T)} \CMI((g\circ f)(A)), (g\circ f)(B))$ as defined in Eq. \ref{eqn: optimal pc g}.

Pixels in $A$ are mapped to either $\bar{x}$ or $\ubar{x}$ by $g_1$. Let $\#A_{\max}$ be the fraction of pixels in $A$ mapped to $\bar{x}$ and let $\#A_{\min}$ be the fraction of pixels mapped to $\ubar{x}$ by $g_1$.
Note that $\#A_{\max} + \#A_{\min} = 1$.
The proportions of pixels in $A$ mapped to $\bar{t}$ and $\ubar{t}$ by $g_2 \circ f$ are also $\#A_{\max}$ and $\#A_{\min}$. To see this is the case, recall that $g_1$ and $g_2$ are constructed as in Eq. \ref{eqn: optimal pc g}, so $g_1(a_i) = \bar{x}$ implies $P_A(a_i) \geq P_B(a_i)$; because $f$ is injective, it follows that $P_{f(A)}(f(a_i)) \geq P_{f(B)}(f(a_i))$  and hence $(g_2 \circ f)(a_i) = \bar{t}.$ Analogously, $g_1(a_i) = \ubar{x}$ implies $(g_2 \circ f)(a_i) = \ubar{t}$.

As above, let $\#B_{\max}$ and $\#B_{\min}$ be the fractions of pixels in $B$ mapped to $\bar{x}$ and $\ubar{x}$ by $g_1$ or, equivalently, to $\bar{t}$ and $\ubar{t}$ by $ g_2 \circ f$. 
Then,
\begin{sizeddisplay}{\footnotesize}
\begin{align*}
    PC_X(A,B) & = \mu[g_1(A)] - \mu[g_1(B)]\\
    &= \bar{x}(\#A_{\max}) + \ubar{x}(\#A_{\min}) - \left (\bar{x}(\#B_{\max}) + \ubar{x}(\#B_{\min}) \right ).
\end{align*}
\end{sizeddisplay}
Similarly, we have $PC_T(A,B) = \bar{t}(\#A_{\max}) + \ubar{t}(\#A_{\min}) - \left (\bar{t}(\#B_{\max}) + \ubar{t}(\#B_{\min}) \right )$. So, $(\bar{x} - \ubar{x}) PC_T(f(A), f(B)) - (\bar{t} - \ubar{t}) PC_X(A,B) = (\bar{x} \ubar{t} - \ubar{x} \bar{t}) (\#A_{\max} + \#A_{\min} - \#B_{\max} - \#B_{\min}) = 0$ because
$\#A_{\max} + \#A_{\min} = \#B_{\max} + \#B_{\min} = 1$, completing the lemma.
\end{proof}
In particular, this lemma shows that PC scales linearly with the range of values in $X$, which is typically defined by the image format. For example, an 8-bit image usually consists of integer values $\{0, 1, \ldots, 255\}$, which is the setting for the original definition of PC. A 16-bit image might have values from $\{0, \ldots, 2^{16}-1\}$, in which case conversion to this format would change the PC by a factor of $\frac{2^{16} - 1}{2^8 -1}.$ Or, if the 16-bit image is restricted to have values between 0 and 1, conversion would change PC by a factor of $1/255.$


\section{Properties \& Equalities for NPC} \label{Sec:PC-Results}
Lemma \ref{lemma: PC scaling with S} shows how PC scales with the range of values given by an image format. An analogous statement for NPC shows its invariance to the image format. 
\begin{lemma} \label{lemma: NPC injective invariance}
    For any injective $f: X \to T \subset \R$, it holds that 
    \begin{equation*}
        NPC(f(A), f(B)) = NPC(A,B).
    \end{equation*}
\end{lemma}
\begin{proof}
    Let $Y$ be the set of values in $A$ or $B$. If $h^{opt}_{A,B} \in \argmax_{h \in H(A,B)} \CMI(h(A), h(B))$ and $h^{opt}_{f(A),f(B)} \in \argmax_{h \in H(f(A),f(B))} \CMI((h \circ f)(A), (h \circ f)(B))$ as defined by Eq. \ref{eqn: optimal npc g}, then $h^{opt}_{A,B}(y) = (h^{opt}_{f(A),f(B)} \circ f)(y)$ for all $y \in Y$. So, $\mu[h^{opt}_{A,B}(A)] = \mu[(h^{opt}_{f(A),f(B)} \circ f)(A)]$ and $\mu[h^{opt}_{A,B}(B)] = \mu[(h^{opt}_{f(A),f(B)} \circ f)(B)]$, which proves the lemma.
\end{proof}
A corollary of Lemma \ref{lemma: PC scaling with S} gives an equivalent formulation for NPC in terms of PC with the appropriate rescaling.
\begin{cor} \label{cor: NPC and PC relation}
    Given an image with values from $X$, it holds that
    \begin{equation*}
        NPC(A,B) = \frac{PC_X(A,B)}{\max(X) - \min(X)}.
    \end{equation*}
\end{cor}
\begin{proof}
    Apply Lemma \ref{lemma: PC scaling with S} with $f(x) = \frac{x - \min(X)}{\max(X) - \min(X)}.$ Because $f$ is injective, $PC_{f(X)}(f(A), f(B)) = \frac{PC_X(A,B)}{\max(X) - \min(X).}$. Since $\min(f(X)) = 0$ and $\max(f(X)) = 1$ we see that $PC_{f(X)}(f(A), f(B)) = NPC(A,B)$.
\end{proof}
This proof shows that NPC is directly equivalent to normalizing an image with $f(x) = \frac{s - \min(X)}{\max(X) - \min(X)}$, and then computing PC. Because of the direct relationship between PC and NPC, many of the observations in \cite{shaus2017-PC-Math} also hold for NPC (e.g. symmetry in arguments $A$ and $B$ or the fact that NPC can be considered an equivalence relation among images). While PC is always in the range $[0, \max(X) - \min(X)]$, NPC has values in $[0,1]$.

Because of its beneficial properties, we use NPC for the remainder of this paper.
However, analogous results for PC can be obtained with the appropriate scaling according to Corollary \ref{cor: NPC and PC relation}.

Next, we prove some equalities for NPC, including its equivalence to the total variation distance \cite{Tsybakov2009_TV} between probability measures, which we denote by $\delta_{tv}(P_A, P_B)$, thinking of $P_A$ and $P_B$ as probability mass functions. In discrete cases like ours, $\delta_{tv}(P_A, P_B)$ can be defined as an $\ell_1$ distance between $P_A$ and $P_B$, i.e.
\begin{align*}
    \delta_{tv}(P_A,P_B) &:= \frac{1}{2} \norms{P_A(x) -P_B(x) }_1\\
    &:= \frac{1}{2} \sum_{x \in X} \norm{P_A(x) - P_B(x)}.
\end{align*}
This equality and our other equalities characterizing $NPC$ are summarized in the following theorem.
\begin{thm} \label{thm: NPC identities}
    Given $A$ and $B$, with associated distributions $P_A$ and $P_B$ defined on some set $X$ (containing all values present in $A$ and $B$), it holds that
    \begin{align}
        NPC(A,B) &= 1 - \sum_{x \in X} \min(P_A(x), P_B(x)) \label{eqn: min identity}\\
        &= \sum_{x \in X} \max(P_A(x), P_B(x)) - 1 \label{eqn: max identity}\\
        &= \frac{1}{2} \norms{P_A(x) -P_B(x) }_1 \label{eqn: l1 identity}\\
        &= \delta_{tv}(P_A,P_B). \label{eqn: TV identity}
    \end{align}
\end{thm}
\begin{proof}
From Eq. \ref{eqn: optimal npc g}, we can write $h^{opt}_{A,B}(x) = \mathbbm{1} [(P_A) (x) \geq (P_B)(x)]$, where $\mathbbm{1}$ is the indicator function. With this notation,
\begin{align*}
        \mu[h^{opt} (A)]
        &= \sum_{x \in X} P_A (x) \mathbbm{1} [P_A(x) \geq P_B(x)],
\end{align*}
and likewise for $\mu[h^{opt}(B)]$. For brevity, we will use the notation $M_{A,B}(x) = \max(P_A(x), P_B(x))$ and $m_{A,B}(x) = \min(P_A(x), P_B(x))$. For any given $x \in X$, whether $P_A(x) \geq P_B(x)$ and $P_A(x) < P_B(x)$, it holds that
\begin{align*}
    (P_A(x)-P_B(x))\mathbbm{1}[P_A(x) \geq P_B(x)] = P_A(x)-m_{A,B}(x).
\end{align*}
Applying this to the definition of NPC yields Eq. \ref{eqn: min identity} as follows:
\begin{sizeddisplay}{\small}
\begin{align*} 
    NPC(A,B) 
    &= \sum_{x \in X} (P_A (x) - P_B(x))\mathbbm{1} [P_A(x) \geq P_B(x)]\\
    &=1 - \sum_{x \in X} m_{A,B}(x).
\end{align*}
\end{sizeddisplay}
Similarly,
\begin{sizeddisplay}{\small}
\begin{align*}
    NPC(A,B) &= 1 - \biggl (\sum_{x \in X} m_{A,B}(x) + \sum_{x \in X} M_{A,B}(x) - \sum_{x \in X} M_{A,B}(x) \biggr)\\
    &=1 - \Bigl ( 2 - \sum_{x \in X} M_{A,B}(x)\Bigr )\\
    &= \sum_{x \in X} M_{A,B}(x) - 1.
\end{align*}
\end{sizeddisplay}
Using the fact that
\begin{equation} \label{eqn: min max dif}
    M_{A,B}(x) - m_{A,B}(x) = |P_A(x) - P_B(x)|,
\end{equation}
we can show that
\begin{sizeddisplay}{\small}
\begin{align*}
   NPC(A,B) &=1 - \sum_{x \in X} m_{A,B}(x) \\ 
    &= 1 - \sum_{x \in X} \biggl ( M_{A,B}(x) - \norm{P_B(x)-P_B(x)} \biggr )\\
    &= 1 + \frac{1}{2}\norms{P_A(x)-P_B(x)}_1 \\
    & \qquad - \sum_{x \in X}  \biggl ( M_{A,B}(x) -\frac{1}{2}\norm{P_A(x)-P_B(x)} \biggr ).
    \end{align*}
\end{sizeddisplay}
To obtain Eq. \ref{eqn: l1 identity}, we prove that the second term in the last equality above is equal to 1 by again applying Eq. \ref{eqn: min max dif}, i.e.
\begin{sizeddisplay}{\small}
\begin{align*}
    1  
    &= \sum_{x \in X}  \Bigl (\frac{1}{2}M_{A,B}(x)+\frac{1}{2}m_{A,B}(x)\Bigr )\\
    &=\sum_{x \in X}  \left ( M_{A,B}(x) -\frac{1}{2}\norm{P_A(x)-P_B(x)} \right ).
\end{align*}
\end{sizeddisplay}
Alternatively Eq. \ref{eqn: l1 identity} also follows from Eq. \ref{eqn: min identity} using Scheffé's Theorem \cite{Tsybakov2009_TV, scheffe1947_equivalence}.

The final equality given by Eq. \ref{eqn: TV identity}, i.e.equivalence to the total variation distance, follows from its definition, taking $P_A$ and $P_B$ to be appropriate probability mass functions.
\end{proof}

Each equality of Theorem \ref{thm: NPC identities} gives insight about NPC. First, Eq. \ref{eqn: min identity} reflects an interpretation of NPC as an accuracy rate. In particular, we can consider $h^{opt}_{A,B}$ from Eq. \ref{eqn: optimal npc g} as a binary classification function. Then, the fraction of pixels in $A$ that are incorrectly classified as belonging to class $B$ is $e_A := \sum_{x \in X} P_A(x) \mathbbm{1} (P_A(x) < P_B(x))$. With a similar expression for $e_B$, the fraction of pixels from $B$ that are misclassified as belonging to $A$, we see that $e_A + e_B = \sum_{x \in X} \min(P_A(x), P_B(x))$. So, the accuracy of our binary classification is $NPC(A,B) = 1 - e_A - e_B$. The conception of PC in this way first appeared in \cite{shaus2017-PC-Math}, where a similar argument described PC in terms of error estimation, with an optimal binarization minimizing the rate of false positives and false negatives ($e_A$ and $e_B$ above). In that setting, the rate was multiplied by $255$ due to the 8-bit images being used, directly corresponding to the scaling between PC and NPC. 

The second equation of Theorem \ref{thm: NPC identities}, Eq. \ref{eqn: max identity} is used later in Section \ref{Sec:Multiple Sources} to generalize NPC to multiple sources.

Third, Eq. \ref{eqn: l1 identity} gives a potentially faster way of computing NPC than the 4-step algorithms of \cite{shaus2017-PC-Math, faigenbaum2012-PC-Method-Ostraca}. We can directly compute (and normalize) the histograms of values for $A$ and $B$ before taking their difference, giving a time complexity of $O(|A| + |B| + |X|)$. This is similar to the time complexity as in \cite{shaus2017-PC-Math} when  $|X|$ is considered constant. However, by not computing the mean $\mu[g^{opt}_{A,B}(A)]$ and $\mu[g^{opt}_{A,B}(B)],$ we avoid iterating over $A$ and $B$ a second time and can reduce computation. 

Finally, the equivalence to total variation in Eq. \ref{eqn: TV identity} is useful beyond proving that NPC (and PC) is a metric. Most importantly, NPC can instead be defined as a total variation distance, which allows us to extend NPC to the continuous case by setting $NPC(A,B) = \delta_{tv}(P_A, P_B)$ when $P_A$ and $P_B$ are defined on continuous domains. The total variation distance is a special kind of integral probability metric, which has the more general structure
\begin{equation*}
D_{\mathcal F} (P_A, P_B) = \sup_{f \in \mathcal{F}} \norm{\mathbb{E}_{Y \sim P_A} f(Y) - \mathbb{E}_{Z \sim P_B} f(Z)}.
\end{equation*}
This difference of means is also in the definition NPC with the difference $\mu[h(A)] - \mu[h(B)]$.
The total variation distance is the integral probability metric with $\mathcal{F} = \{f: X \to \{0, 1\}\}$, mirroring the functions contained in $H(A,B)$.


\section{Multi-Class Normalized Potential Contrast}\label{Sec:Multiple Sources}

\begin{figure}[t]
\centering
\includegraphics[width=0.24\textwidth]{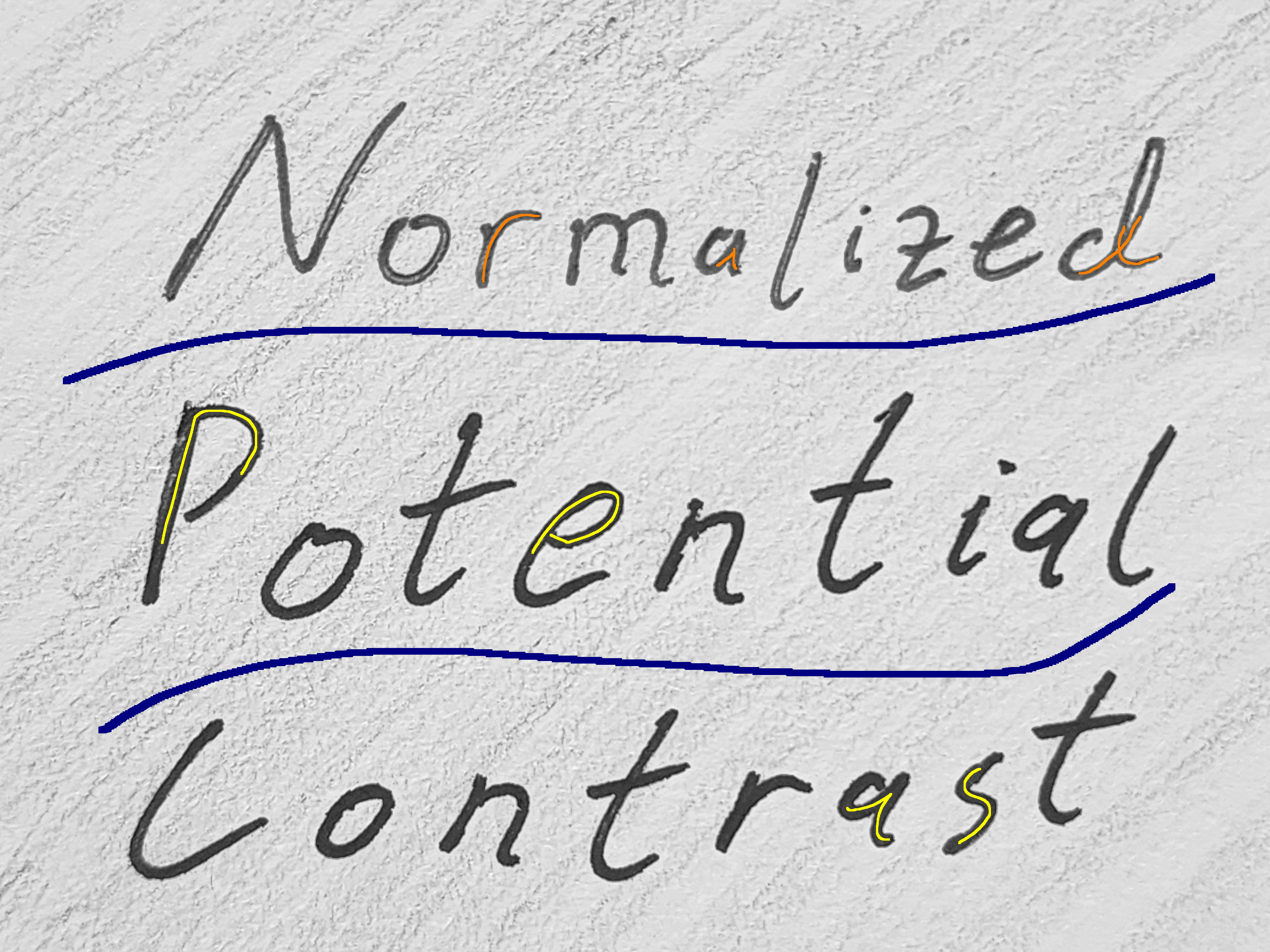}
\includegraphics[width=0.24\textwidth]{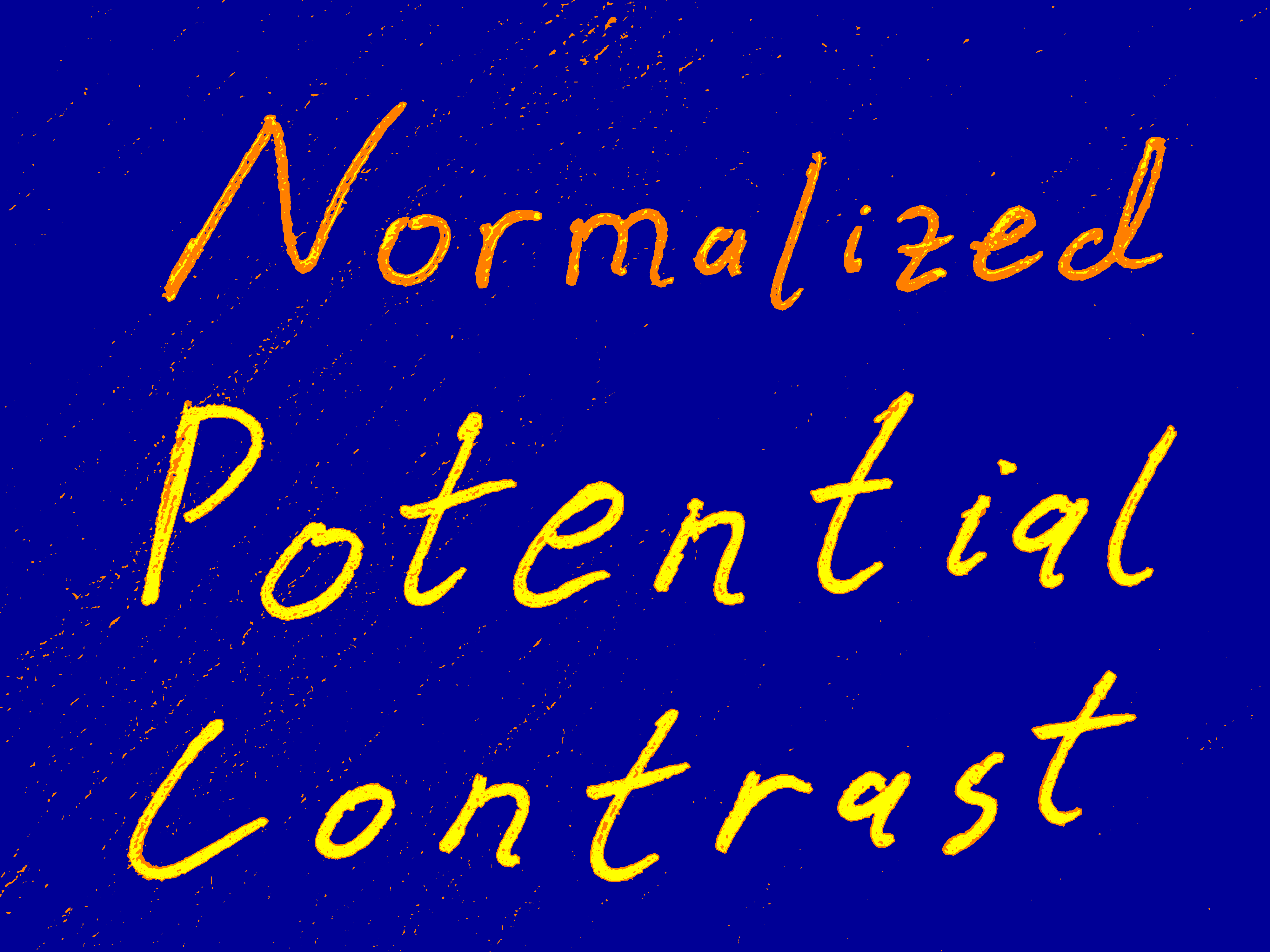}
\caption{Left: The example image from Figure \ref{fig: normalized potential contrast}, now with three different classes labeled. Labels for one ink are in yellow, for a second ink are in orange, and for the background are in blue. Top Right: a multi-class NPC result, where the colors reflect the class segmentation. The multi-class NPC has a value of 0.885.}
\label{fig: multi-class npc}
\end{figure}
A generalized multi-class NPC allows us to compute NPC when more than two classes are present. This could arise in various contexts, e.g. manuscript bleedthrough, palimpsest, or many different inks. Given $n$ classes, NPC could be applied to all $\binom{n}{2}$ pairs, yielding $\binom{n}{2}$ values and binarizations. Alternatively, for cases where interpreting all of these relationships simultaneously is of interest, we extend NPC to multiple classes by generalizing  Eq. \ref{eqn: max identity} from Theorem \ref{thm: NPC identities}. 

Given $n$ classes, we denote $A_i= (a_1, \ldots, a_{m_i})$ as labeled samples for a class $i \in \{1, \ldots, n \}$, each with a discrete distribution $P_{A_i}$.
\begin{definition}
We define the multi-class NPC of $A_1, \ldots A_n$ as
    \begin{sizeddisplay}{\small}
    
    \begin{equation} \label{eqn: multi class max}
        NPC(A_1, \ldots, A_n) := \frac{ -1 + \sum_{x \in X} \max_{1 \leq i \leq n} P_{A_i}(x)}{n-1}.
    \end{equation}
    \end{sizeddisplay}
\end{definition}
The scaling ensures that $NPC(A_1, \ldots, A_n) \in [0,1]$.
For pixel-wise segmentation, we assign a pixel with value $x$ to the class $A_{i}$ when $i \in \argmax_{1 \leq i \leq n} P_{A_i}(x)$, i.e. to the class that has the highest fraction of its sampled pixels at that value $x$. An example of such a segmentation is given in Figure \ref{fig: multi-class npc}.
    We also note that  multi-class PC can be defined by scaling multi-class NPC like in the two-class case:
    \begin{equation*}
    PC(A_1, \ldots,A_n):= (\max(X) - \min(X)) \cdot NPC(A_1, \ldots , A_n).    
    \end{equation*}
We can now state an analog of Theorem \ref{thm: NPC identities} for multi-class NPC.
\begin{thm}
    Suppose we have $n$ classes with sampled pixels $A_i$ and distributions $P_{A_i}$ defined on $X$. For a given $x \in X$, let $P_{A_x^{(i)}}$ be a reordering of $P_{A_i}$ from smallest to largest, i.e.  $\min_{1 \leq i \leq n} P_{A_i}(x) = P_{A_x^{(1)}}(x) \leq P_{A_x^{(2)}} (x)\leq \ldots \leq P_{A_x^{(n)}}(x) = \max_{1 \leq i \leq n} P_{A_i} (x)$. Then,
    \begin{sizeddisplay}{\small}
    \begin{align}
        NPC(A_1, \ldots, A_n) 
        &= 1 - \frac{1}{n-1}\sum_{x \in X} \sum_{i=1}^{n-1} P_{A_{x}^{(i)}}(x) \label{eqn: multi-class min} \\
        & =
        \frac{1}{n(n-1)} \sum_{x \in X} \sum_{i=1}^{n-1} (P_{A_{x}^{(n)}}(x) - P_{A_{x}^{(i)}}(x) ). \label{eqn: multi-class l1}
    \end{align}
    \end{sizeddisplay}
\end{thm}
The proof is similar to Theorem \ref{thm: NPC identities}. We note that Eqs. \ref{eqn: multi class max}, \ref{eqn: multi-class min}, and \ref{eqn: multi-class l1} are analogs of Eqs. \ref{eqn: max identity}, \ref{eqn: min identity}, and \ref{eqn: l1 identity} respectively.

Additionally, we can rewrite Eq. \ref{eqn: multi-class l1} as 
\begin{sizeddisplay}{\footnotesize}
\begin{equation*}
    NPC(A_1, \ldots A_n)
    = 
    \frac{1}{n} \sum_{x \in X} \left ( P_{A_{x}^{(n)}}(x) - \frac{1}{n-1} \sum_{i=1}^{n-1} P_{A_{x}^{(i)}}(x) \right ).
\end{equation*}
\end{sizeddisplay}
We can interpret this as the mean, across each possible value $x \in X$, of the difference between the distribution with largest fraction of pixels that have value $x$ and the average of the remaining distributions.

We can also interpret multi-class NPC as an accuracy rate similarly to Section \ref{Sec:PC-Results}. In particular, if $e_i$ gives the proportion of pixels from $A_i$ that are misclassified as belonging to a different class $j \neq i$, then 
\begin{equation*}
    NPC(A_1, \ldots, A_n) = 1 - \frac{1}{n-1}\sum_{i=1}^n e_i.
\end{equation*}
This follows from Eq. \ref{eqn: multi-class min}, since the sum $\sum_{i=1}^{n-1}P_{A_x^{(i)}}(x)$ gives an error rate across all classes for a pixel value $x$ because $x$ is assigned to class $i$ when $P_{A_i}(x)$ is maximal (among all distributions evaluated at $x$).

\section{Application} \label{Sec:Application}

\begin{figure}[t!]
\centering
\frame{\includegraphics[width=0.48\textwidth]{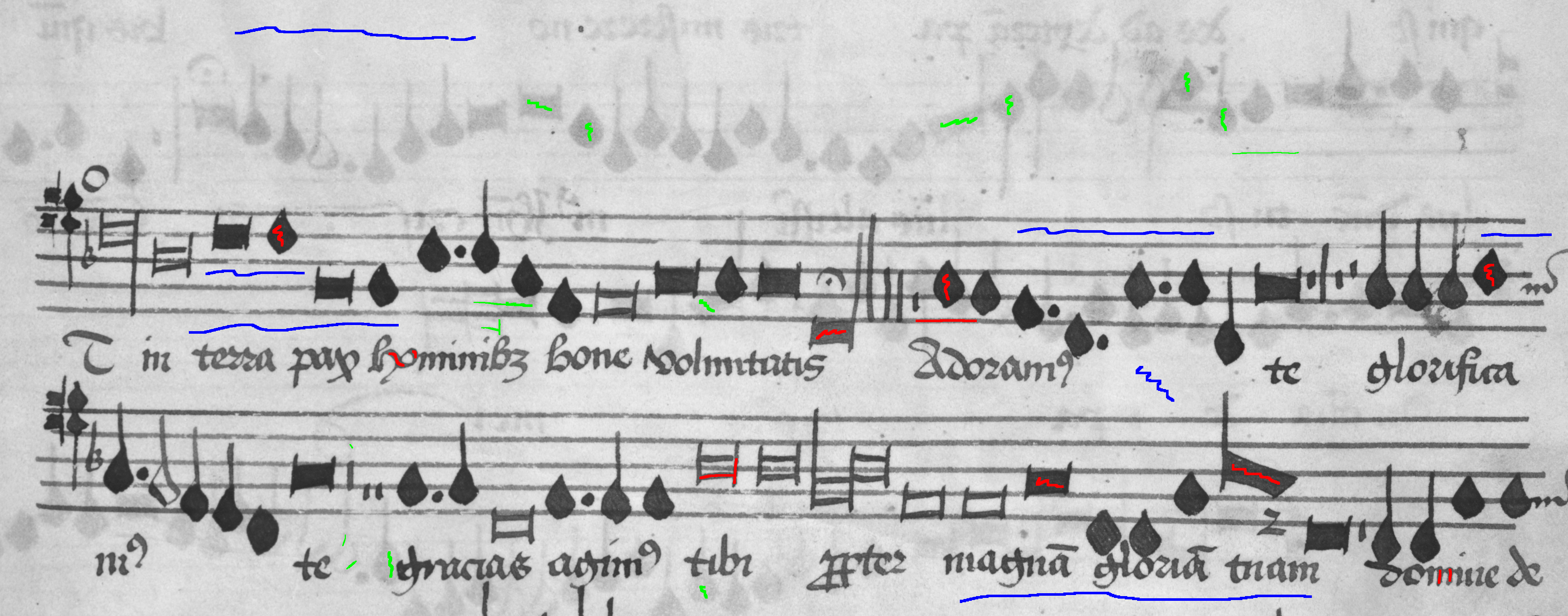}}
\frame{\includegraphics[width=0.48\textwidth]{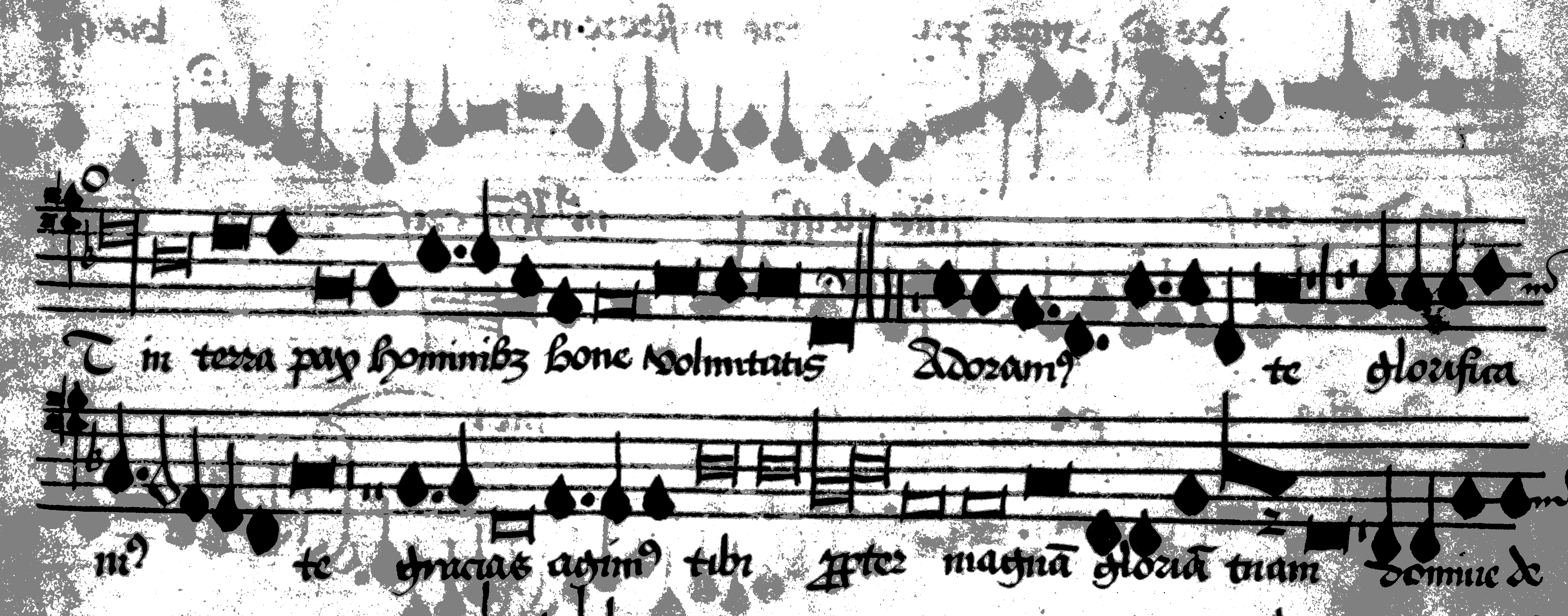}}
\frame{\includegraphics[width=0.48\textwidth]{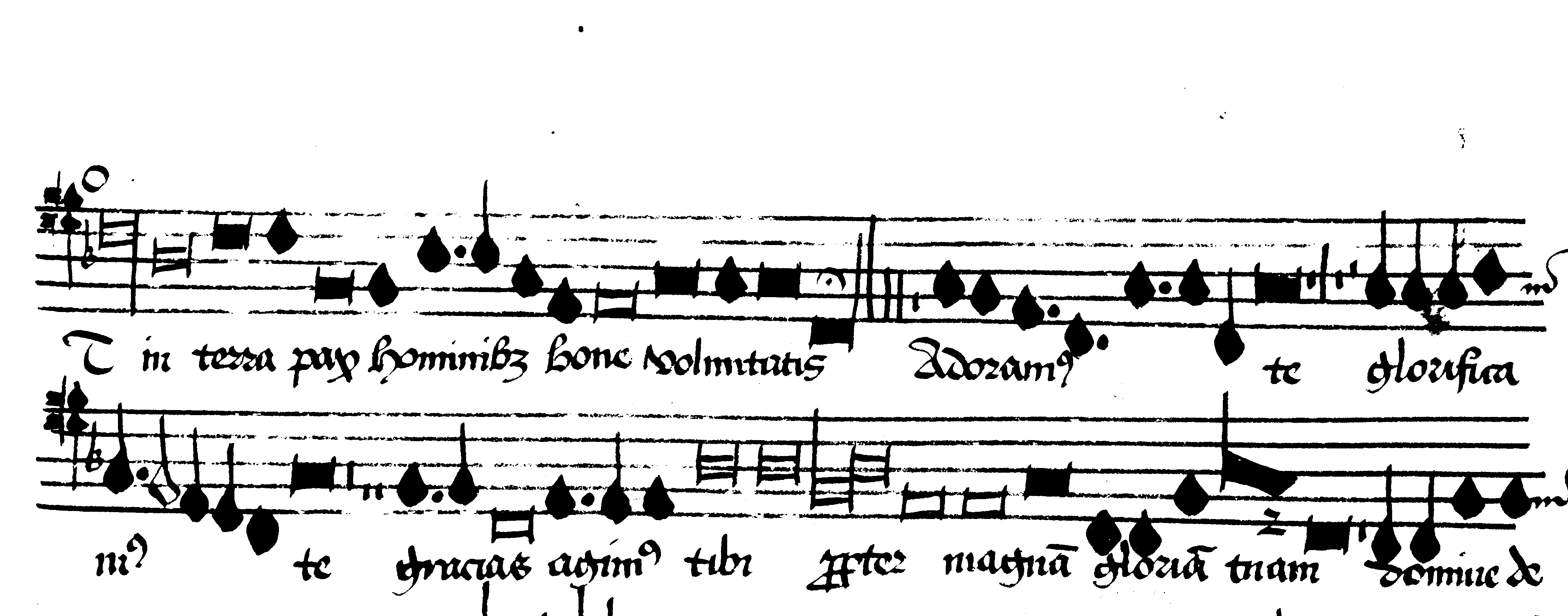}}
\caption{(Top) A grayscale image showing an excerpt from GB-Cgc 667/760, page 19, with foreground notation (red), bleedthrough (green), and background (blue) labeled. (Middle) A three-class NPC result with foreground in black, bleedthrough in gray, and background in white. The three-class NPC value is 0.965. The two-class NPC between foreground notation and background is nearly 1, the two-class NPC between foreground and bleedthrough is 0.996, and the two-class NPC between bleedthrough and background is 0.934. (Bottom) The three-class NPC segmentation showing only foreground notation.\
Original imaging by DIAMM (\url{diamm.ac.uk}).
Image use by kind permission of the Master and Fellows of Gonville and Caius College, Cambridge.}
\label{Fig: Manuscript Example}
\end{figure}

To illustrate the utility of multi-class NPC, we apply it to photographed page from the Caius Choirbook (GB-CGC 667/760), a music manuscript dated to the late 1520s and currently at Gonville and Caius College, Cambridge \cite{DIAMM-webpage}. Like many historical manuscripts, this choirbook exhibits substantial bleedthrough, showing ink from the back of the page as well as the front. We apply NPC with three classes: written notation on the open page, bleedthrough, and background. The result, shown in Figure \ref{Fig: Manuscript Example}, is a strong multi-class NPC value, suggesting that the labeled pixels are easily separated based only on their intensity values. This is reflected in the segmentation, where foreground text is easily isolated, allowing us to remove the bleedthrough from the page and visualize the music without undesired noise from the background.

The two-class NPC values also reflect what we expect: the foregrorund is easy to separate from background and from bleedthrough, but separating background from bleedthrough is more difficult. However, for many applications a single value is more useful, one which accounts for all classes rather than only pairwise comparisons.

The imperfections in Figure \ref{Fig: Manuscript Example} also exemplify some of the challenges of using NPC. For example, because NPC is a global measure and relies only on the histogram of grayscale values (discarding spatial information), changes in brightness across a page, such as those from uneven lighting or dirt, can result in a lower multi-class NPC when that may not be desired or expected.

Finally, the NPC value also heavily depends on the labeled pixels. While  labeling can take time, usually only a small number of pixels are needed from each class. This labeling also allows a user to adapt the measure to a particular task.



\section{Conclusion} \label{Sec:Conclusion}
Potential contrast is an image contrast and quality measure that relies on task-dependent labels. We introduce a scaled version, normalized potential contrast (NPC), that is commensurate across image formats and applications, which retains many of the desirable properties of potential contrast.

We prove equalities that support different interpretations of NPC and show it is equivalent to the total variation distance on the probability distributions of sampled pixels, allowing generalization to continuous domains. One of the other equivalences allows us to define NPC when there are more than two classes, namely multi-class NPC.

Such generalizations enable applications to a much broader array of contexts, including where potential contrast has already found success. We showcase an example with three classes in a music manuscript exhibiting bleedthrough,  demonstrating how multi-class NPC can provide a valuable, interpretable, and task-adaptive image quality and contrast measure in cultural heritage and beyond.


\bibliographystyle{ieeetr}
\bibliography{sources}

\begin{thebibliography}{10}

\bibitem{shaus2017-PC-Math}
A.~Shaus, S.~Faigenbaum-Golovin, B.~Sober, and E.~Turkel, ``Potential contrast--a new image quality measure,'' {\em Electronic Imaging}, vol.~29, pp.~52--58, 2017.

\bibitem{faigenbaum2012-PC-Method-Ostraca}
S.~Faigenbaum, B.~Sober, A.~Shaus, M.~Moinester, E.~Piasetzky, G.~Bearman, M.~Cordonsky, and I.~Finkelstein, ``Multispectral images of ostraca: acquisition and analysis,'' {\em Journal of Archaeological Science}, vol.~39, no.~12, pp.~3581--3590, 2012.

\bibitem{sober2014-PC-Method-Ostraca-Conference}
B.~Sober, S.~Faigenbaum, I.~Beit-Arieh, I.~Finkelstein, M.~Moinester, E.~Piasetzky, and A.~Shaus, ``Multispectral imaging as a tool for enhancing the reading of ostraca,'' {\em Palestine Exploration Quarterly}, vol.~146, no.~3, pp.~185--197, 2014.

\bibitem{faigenbaum2022-PC-Example_BITS}
S.~Faigenbaum-Golovin, A.~Shaus, and B.~Sober, ``Computational handwriting analysis of ancient hebrew inscriptions—a survey,'' {\em IEEE BITS the Information Theory Magazine}, vol.~2, no.~1, pp.~90--101, 2022.

\bibitem{faigenbaum2017-PC-Example_Biblical}
S.~Faigenbaum-Golovin, A.~Mendel-Geberovich, A.~Shaus, B.~Sober, M.~Cordonsky, D.~Levin, M.~Moinester, B.~Sass, E.~Turkel, E.~Piasetzky, {\em et~al.}, ``Multispectral imaging reveals biblical-period inscription unnoticed for half a century,'' {\em PLoS One}, vol.~12, no.~6, p.~e0178400, 2017.

\bibitem{faigenbaum2014-PC-Example_Hieratic}
S.~Faigenbaum, B.~Sober, I.~Finkelstein, M.~Moinester, E.~Piasetzky, A.~Shaus, and M.~Cordonsky, ``Multispectral imaging of two hieratic inscriptions from qubur el-walaydah,'' {\em {\"A}gypten und Levante/Egypt and the Levant}, pp.~349--353, 2014.

\bibitem{fechner1948_psychology}
G.~T. Fechner, ``Elements of psychophysics, 1860.,'' {\em Readings in the history of psychology.}, pp.~206--213, 1948.
\newblock Place: East Norwalk, CT, US Publisher: Appleton-Century-Crofts.

\bibitem{Tsybakov2009_TV}
A.~B. Tsybakov, {\em Nonparametric estimators}.
\newblock New York, NY: Springer New York, 2009.

\bibitem{Shaus2012-CMI}
A.~Shaus, E.~Turkel, and E.~Piasetzky, ``Quality evaluation of facsimiles of hebrew first temple period inscriptions,'' in {\em 2012 10th IAPR International Workshop on Document Analysis Systems}, pp.~170--174, 2012.

\bibitem{scheffe1947_equivalence}
H.~Scheff{\'e}, ``A useful convergence theorem for probability distributions,'' {\em The Annals of Mathematical Statistics}, vol.~18, no.~3, pp.~434--438, 1947.

\bibitem{DIAMM-webpage}
{Digitial Image Archive of Medieval Music}, ``Gb-cgc ms 667/760 (caius choirbook).'' \url{https://www.diamm.ac.uk/sources/225/}.
\newblock Accessed: 2025-03-13.

\end{thebibliography}

\end{document}